\documentclass[final,1p,times,authoryear]{elsarticle}

\usepackage{amssymb}
\usepackage{pdflscape}
\usepackage{url,hyperref,lineno,microtype,caption,subcaption}
\usepackage{xurl}
\usepackage[onehalfspacing]{setspace}
\usepackage{orcidlink}
\usepackage{multirow}
\usepackage{longtable}%
\usepackage{tabularray}
\usepackage{graphicx}%
\usepackage[normalem]{ulem}
\useunder{\uline}{\ul}{}
\usepackage{textcomp,siunitx}
\usepackage{pdfpages}
\usepackage{enumitem}
\setitemize{noitemsep,topsep=0pt,parsep=0pt,partopsep=0pt}

\journal{Computers \& Education}

\begin{document}

\begin{frontmatter}



\title{\textit{Microtopia}: Exploring the Impact of Interdisciplinary Projects on Ethnic Minority Female Pupils’ Perceptions of Computer Science}

\author[Brunel]{Nadine Aburumman \orcidlink{0000-0003-4578-8738}}
\affiliation[Brunel]{organization={Department of Computer Science, Brunel University of London},
            addressline={Kingston Ln}, 
            city={Uxbridge},
            postcode={UB8 3PH}, 
            country={UK}}

\author[NationalCentralUniversity]{Ju-Ling Shih \orcidlink{0000-0003-2179-9117}}
\affiliation[NationalCentralUniversity]{organization={Graduate Institute of Network Learning Technology, National Central University},
            addressline={No. 300, Zhongda Rd, Zhongli District}, 
            city={Taoyuan City},
            postcode={320}, 
            country={Taiwan}}
            
\author[Brunel]{Cigdem Sengul \orcidlink{0000-0002-6011-9690}}
\author[Brunel]{Monica Pereira \orcidlink{0000-0003-2583-4522}}

\begin{abstract}
This paper presents \textit{Microtopia}, an interdisciplinary programme designed to broaden participation in computer science (CS) among ethnic minority girls. The programme combined coding with design thinking activities, incorporating Artificial Intelligence (AI), the Internet of Things (IoT), and Robotics as key technologies. Learning activities were formulated around the UN Sustainable Development Goals and the Chinese Five Elements philosophy to support problem-based learning. Pupils were organised into "nations" and engaged in sector-based projects (e.g., healthcare, transportation, fashion, tourism, food, architecture). Using pre- and post-questionnaires, we investigated how socioeconomic and ethnocultural factors influenced pupils' preconceptions of CS, and whether participation in \textit{Microtopia} shifted their perceptions. Through statistical analysis of the questionnaire data, we identified significant increases in students’ confidence, enjoyment, and motivation, particularly when computing was presented as relevant to sustainability and global challenges.
\end{abstract}


\begin{highlights}
\item We propose the \textit{Microtopia} programme, an interdisciplinary and collaborative computer science initiative for secondary school pupils that integrates Artificial Intelligence, the Internet of Things, and Robotics computing projects.

\item Our findings from running the \textit{Microtopia} programme with 90 pupils demonstrate positive outcomes, including increased confidence, enjoyment, interest, and willingness to learn more, although computer science is still perceived as difficult.

\item Linking \textit{Microtopia}’s content to real-world sustainability and global challenges enhanced pupils’ engagement, as they took purposeful roles that reinforced the relevance of computer science learning.

\item Our statistical analysis identifies underlying factors shaping pupils’ perceptions of computer science. Pupils from higher socioeconomic status (SES) backgrounds reported greater confidence, whereas those from lower SES backgrounds more often perceived the field as male-dominated.
\end{highlights}
\end{frontmatter}
\section{Introduction}
Women are enrolling in universities and graduating at higher rates than men worldwide. However, their representation in Science, Technology, Engineering, and Mathematics (STEM) fields, particularly computer science (CS), engineering, and physics, remains significantly lower \citep{Klawe2013-increasing, hammond2020-equality}. In the UK, research from 2020 shows 65\% of the STEM workforce is composed of white men, with only 4\% being ethnic minority women \citep{BSA2020}, and fewer than 1\% of IT specialists are Black women. Despite recent increases in female enrolment in computing programmes, gender and racial disparities persist, limiting the diversity of the CS field \citep{BCS2024}.

A growing body of research highlights the complex barriers contributing to this underrepresentation. Stereotypes, lack of early exposure, and masculine computing cultures discourage ethnic minority female students from considering CS careers \citep{wong16-good}. Many women perceive CS as unwelcoming due to gendered expectations, lower self-efficacy, and a lack of role models. Intersectionality further compounds these challenges, as racial and ethnic minority women experience exclusion based on both gender and ethnicity, making it even harder to develop a strong CS identity \citep{williams23-blackwomen}.

To address these disparities, education initiatives must go beyond traditional pedagogies and provide meaningful, interdisciplinary learning experiences \cite{Inclusive2025}. To this end, we designed the \textit{Microtopia} programme to engage ethnic minority female students by integrating coding-based computing projects and design thinking activities linked with real-world industrial applications. Through activities incorporating Artificial Intelligence (AI), the Internet of Things (IoT), and Robotics, the programme intends to foster collaboration, creativity, and a sense of purpose. \textit{Microtopia} seeks to make computer science more relatable and appealing to young learners by aligning projects with themes that reflect the United Nations Sustainable Development Goals (UN SDGs) \citep{UNSDG}.

This study investigates the impact of \textit{Microtopia} on ethnic minority female pupils’ perceptions of CS. In particular, we explore how socioeconomic and ethnocultural factors shape their preconceptions and whether \textit{Microtopia}, an interdisciplinary and collaborative computing programme, can shift their attitudes towards computer science. The programme was delivered through a series of one-day workshops, designed to create meaningful connections between computing and real-world challenges, and also examines how this influences student engagement and motivation in CS. Our research questions are as follows.
\begin{itemize}
\item RQ1: To what extent do socioeconomic factors influence ethnic minority girls’ preconceptions of CS?
\item RQ2: To what extent do ethnocultural factors influence ethnic minority girls’ preconceptions of CS?
\item RQ3: To what extent does \textit{Microtopia} impact ethnic minority girls’ perceptions of CS?
\item RQ4: To what extent does \textit{Microtopia} enable ethnic minority girls to establish meaningful connections between CS and real-world challenges?
\end{itemize}
By addressing these research questions, this paper contributes to ongoing efforts to close the gender and racial gap in computing. The findings offer insights into how innovative, interdisciplinary education strategies can help challenge stereotypes, build confidence, and foster engagement and interest among underrepresented students in computer science.
\section{Literature Review}
\label{Background}
The first part of this section reviews existing literature on the factors contributing to women's underrepresentation in STEM. It then explores the impact of intersectionality, specifically race, ethnicity, and gender. Finally, this section presents various initiatives designed to increase women’s participation in STEM and their effectiveness in fostering a more inclusive environment.

\subsection{Explanations for Underrepresentation of Women in STEM}
The reasons for the underrepresentation of women in STEM fields are believed to be complex. There is also a much-debated paradox: as a country’s income rises, the gender gap for the likelihood of studying in a STEM field widens, and the gap increases with higher measures of national gender equality \citep{hammond2020-equality}. While experts disagree on the reasons behind this phenomenon, one explanation considers the influence of the freedom of choice among women and the lower opportunity cost of forgoing more profitable STEM careers in these countries. Furthermore, nations with high overall gender equity may nonetheless have strong gender-science stereotypes if men dominate science fields, affecting women’s participation in science \citep{Miller2015-representation}.

These stereotype threats play a role in undermining women's interest and performance in STEM \citep{Shapiro2012-stereotype}. Considering computer science as the focus of this study, three key factors are often cited as reasons contributing to why women do not major in CS \citep{Klawe2013-increasing}: 1) They do not find CS interesting. 2) They do not believe they will do well in CS. 3) They feel uncomfortable in computing culture.

Several studies, e.g., \citep{Fisher1997, abou2013proposed, monica2017-exploring, Rezwana2023}, emphasise the importance of prior experience for women's recruitment and retention in computing. In contrast, \citet{monica2017-exploring} discusses prior experience, surprisingly, does not appear to play a role in increasing the belongingness of female students in their study. Therefore, more research is needed to understand the factors that affect recruitment and retention of female students in CS.

\subsection{Intersecting Inequalities: Gender, Race and Ethnicity and Belonging in Computing}
\label{ss:intersections_belonging}
\emph{Intersectionality} considers identities like race, gender, class, and similar categories to be socially constructed, and best understood together rather than in isolation. The notion does not relate only to group membership, but also to the associated vulnerability to discrimination and exclusion due to that membership \citep{ireland18-unhidden}. Only limited research exists on women from different ethnic groups in CS, specifically their conceptions of the field.

Collecting data from museum visits of three science classes from two London schools, observed that even when the girls from racialised minorities displayed behaviours that could be considered in line with successful science student positions (e.g., confident displays of scientific expertise or assertively using interactive exhibits), they were not recognised as such in \citep{dawson20-selfies}. This issue has also been reported in \citep{bloodhart20-outperforming}, where girls are underestimated even when outperforming boys in their studies.

Studies describe that ``it's tough'' to be a woman in computing when there are rampant misperceptions and stereotypes about their academic and intellectual abilities \citep{charleston14-underrepresented, ireland18-unhidden, willis24-black}. Even when girls are considered high-achieving, their association with a science degree can be affected by various social, cultural, and structural factors, as in the case of two 13-year-old working-class British Asian girls, who are the focus of the study carried out by \citet{wong12-identifying}. ``Invisibility'' \citep{ ireland18-unhidden, dawson20-selfies, williams23-blackwomen, fisk24-retaining} and ``isolation'' \citep{charleston14-underrepresented, morton18-blackgirlmagic, rankin21-black,willis24-black} are reported to shape Black women's identification and connections with CS. 

\subsection{Interventions with Potential to Change Intentions and Behaviour Toward 	Computer Science}
Interventions can be broadly categorised with three factors: activities for increasing exposure, activities for creating meaningful work, and activities for improving collaboration. We discuss the literature on these interventions in the following three sections.

\subsubsection{Activities for Increasing Exposure}
Extracurricular activities for both boys and girls, such as museum visits, clubs, and robotics and coding camps
are found promising for building interest, changing attitudes, and boosting confidence and self-efficacy \citep{kesar18-STEMgap, hammond2020-equality,blake24-future}. These activities are particularly important for those who lack opportunities to engage in everyday science experiences at home \citep{ireland18-unhidden}. Similarly, Harvey Mudd College in the United States achieved significant success in increasing female participation in computing \citep{Klawe2013-increasing}. As part of this intervention, female students were given the opportunity to meet role models at the Grace Hopper Celebration of Women in Computing, helping them to envision who they could become \citep{Klawe2013-increasing}.

\subsubsection{Activities for Creating Meaningful Work}
Constructivist pedagogies fit well with women's desire to engage in ``computing with a purpose''. Knowledge-building processes are facilitated by constructing tangible, shareable artefacts such as stories, models, and computer programs, often in a social context \citep{papert91-constructionism}. \citet{carbonaro10-game} used interactive game adventure authoring as an enjoyable introduction to computing and find that this activity is gender-neutral, placing males and females on equal footing as computer game builders, as opposed to computer game playing, which favours males \citep{volman05-genderethnic}. \citet{carvalho20-gendergap} also proposed serious games as part of the CODING4GIRLS initiative to promote the development of programming skills among girls and found this learning methodology captivating for both girls and boys \citep{coding4girls}. \citet{Krohn2020-gendergap} redesigned programming courses using COOL informatics \cite{sabitzer13-cool} and brain-based programming \citep{sabitzer14-brainbased}. This teaching methodology, which incorporates a conversational, cooperative style into hands-on learning-by-doing, was also found as or more agreeable to girls \citep{zastavker06-wie}.

\subsection{Activities for Improving Collaboration} 
Creating collaborative environments was also central to the intervention described by Escobar et al. \citep{escobar2021engaging}. Their programme fostered a peer learning community, where Black women engaged in culturally responsive, project-based activities and received ongoing support from teachers and near-peer mentors. Furthermore, groupwork-based collaboration is found beneficial for both men and women, where there is a positive correlation between student participation in small group work and their engagement, motivation, satisfaction, and understanding. Still, the method remains controversial as women report higher levels of stress and frustration in mixed-gender groups \citep{zastavker06-wie}, which may be due to perceiving peers as unsupportive and competitive. 

However, not all interventions listed above are helpful for all groups \citep{fisk24-retaining}. While significant predictors of white women's computing persistence include confidence building and working with others who are like themselves, \cite{fisk24-retaining} reasoned that women pursuing computing must already be confident and committed and, hence, may not benefit from these interventions. Given that their analysis explores women who are already interested in computing, different interventions may be needed to attract women to the field of computing. In this paper, we propose the \textit{Microtopia} programme, which aims to influence ethnic minority young women's intentions and behaviours by fostering interest in computer science and shifting their perceptions of the field. We hypothesise that providing female high school students with opportunities to participate in interdisciplinary, team-based projects aligned with real-world problems can positively shape their beliefs towards CS.
\section{\textit{Microtopia} Programme}
\label{Method}
\subsection{Instructional Design}
\begin{figure}[h!]
\centering
\includegraphics[width=1\linewidth]{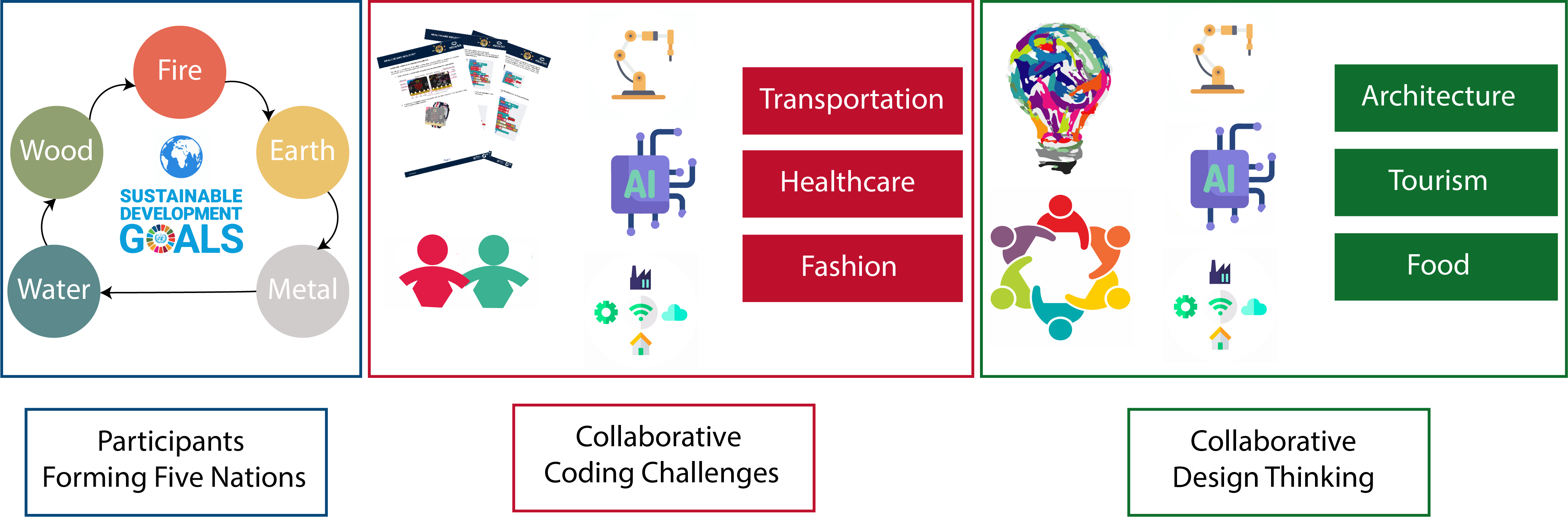}
\caption{An overview of the activity framework used in the design of the \textit{Microtopia} programme.}
\label{fig:WSFM}
\end{figure}

 The \textit{Microtopia} workshops used insights from constructive pedagogies and incorporated a game-based approach to help students develop essential technical skills applicable across various fields.  To this end, Microtopia used the following gaming elements and mechanisms to engage students and facilitate learning (see Figure \ref{fig:WSFM}).  
\begin{itemize}
\item Collaboration: Students work in groups to complete computing problems using their communication and problem-solving skills \citep{Shahin2022}.
\item Role-Playing: Students take on roles in their groups, each with its characteristics and goals, fostering purpose and understanding of diverse perspectives \citep{lim2011technology}. 
\item Competition: Competition across groups encourages students to develop technical problems within the given time frame \citep{juvonen2019promoting}. 
\item Ranking: Each group's performance is evaluated and ranked accordingly, fostering healthy competition and motivation among students to excel in their development efforts \citep{Werner2014}.
\end{itemize}
These elements were implemented in \emph{Microtopia}, which was designed around real-world problems aligned with the SDGs and framed within Chinese philosophy as the underpinning approach \cite{wang2012yinyang}. Students took on group roles as different nations, working to solve industry-specific challenges. In the Collaborative Coding Challenges stage, students worked in groups to complete three coding tasks related to their assigned industry. In the Collaborative Design Thinking stage, the groups were then guided to develop innovative solutions to problems drawn from a different industry.

\subsubsection{Participants Forming Five Nations}
In \emph{Microtopia}, students were grouped into five nations, Metal, Wood, Water, Fire, and Earth, named after the Chinese Five Elements, or Wu Xing. These elements hold deep symbolic and philosophical significance in Chinese culture, emphasising the dynamic interactions needed to maintain balance in life and the environment. This theme connects naturally to the UN SDGs through its underlying interpretations.
\begin{figure}[h!]
\centering
\includegraphics[width=1\linewidth]{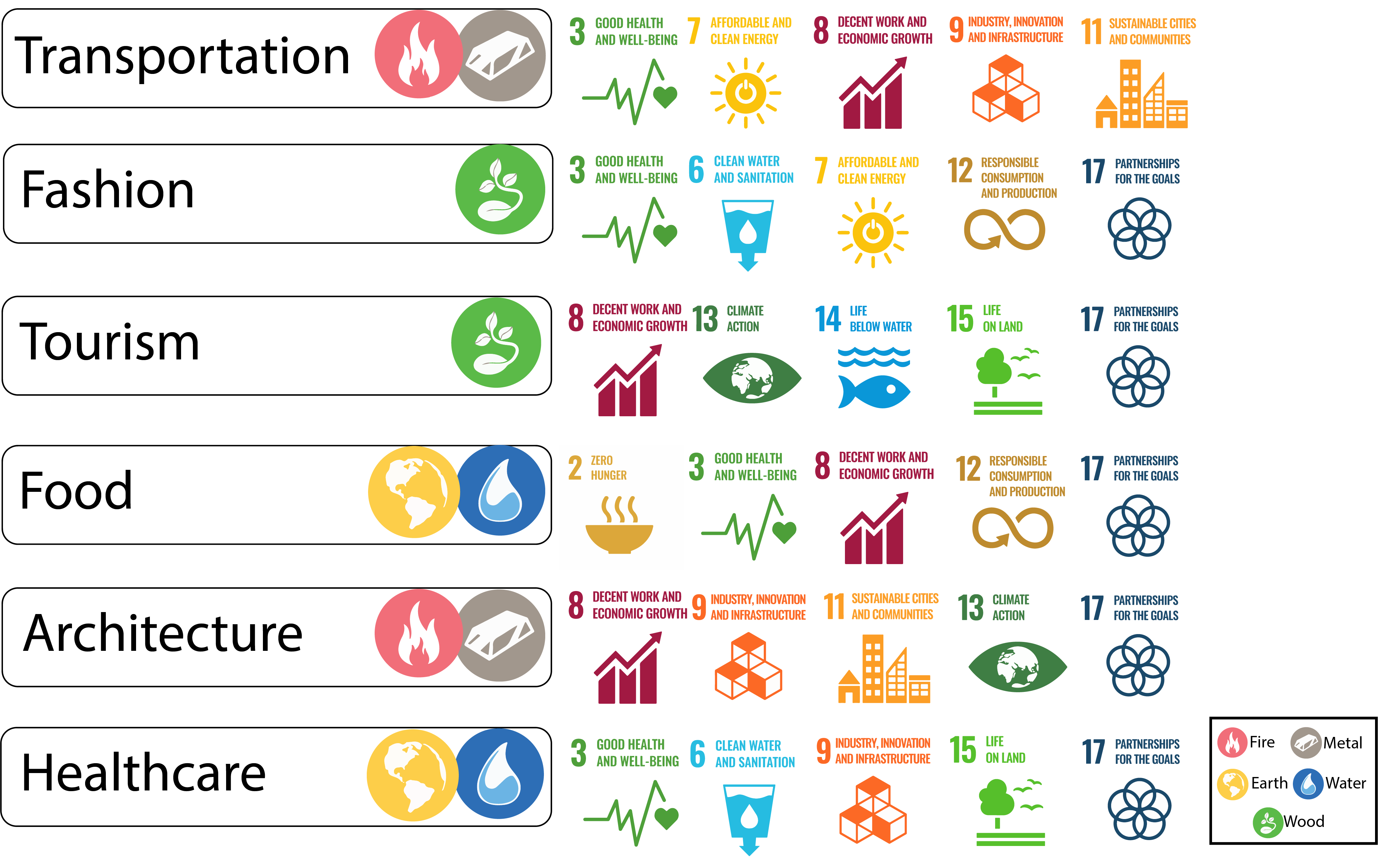}
\caption{Each nation, named after one of the Chinese Five Elements, is tasked with challenges spanning different industries and Sustainable Development Goals.}
\label{fig:MUNSG}
\end{figure}

The nations reflect different aspects of nature and life: Metal symbolises resilience and innovation, representing renewable energy and sustainable materials. Wood reflects growth and regeneration, highlighting forests’ role in climate change mitigation and biodiversity. Water signifies balance and purification, emphasising conservation and clean water access. Fire represents energy and transformation, supporting the shift to renewable energy and sustainable urban planning. Earth symbolises stability and nourishment, stressing sustainable land use, soil conservation, and food security. Together, these elements provide a framework for building sustainable and thriving nations. Each nation is assigned two industries: One of the transportation, fashion, healthcare industries, and one of the tourism, food technology, and architecture industries. This choice of industries was selected to reflect the characteristics of their respective elements (see Figure \ref{fig:MUNSG}). The groups were introduced to their assigned nation and the technologies (IoT, AI, and Robotics) and were tasked to employ these technologies to advance their nation’s development.
\subsubsection{Collaborative Coding Challenges}
\begin{figure}[h!]
\centering
\includegraphics[width=1\linewidth]{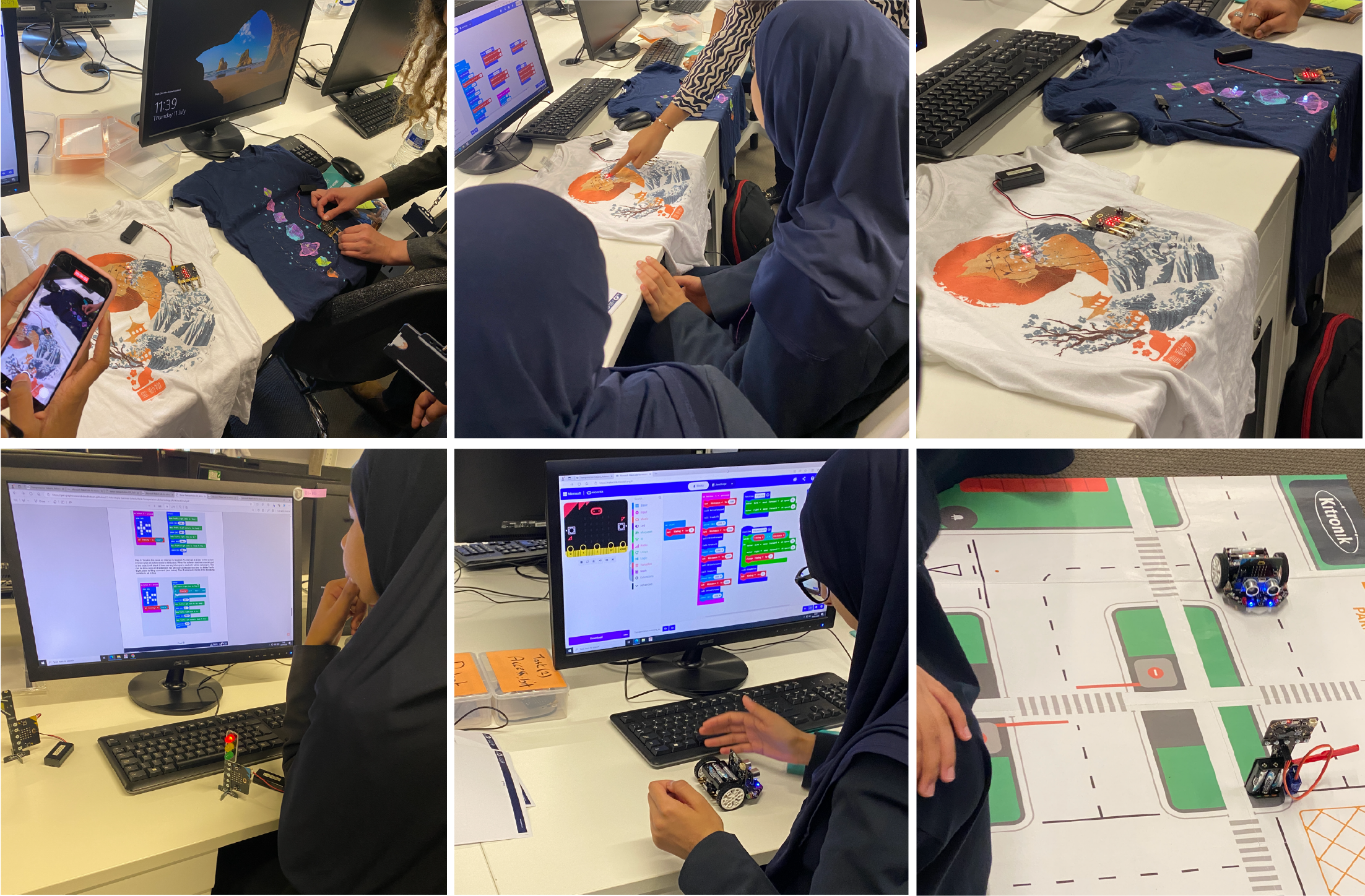}
\caption{Students engage in coding-based challenges using BBC micro:bits and other materials.}
\label{fig:MICROBIT}
\end{figure}

In this stage, groups are given 3 hours to collaborate to solve coding-based challenges themed around IoT, AI, and Robotics, the solutions of which reflect valuable contributions to sustainability in their assigned industries: transportation, fashion, and healthcare. Worksheets were provided for each computing challenge, following \citet{Yadav2021}, to promote active and collaborative learning while introducing fundamental programming concepts (see Supplementary Material for a sample of the coding worksheets). These worksheets also serve as instructional tools, incorporating targeted questions and contextual information to guide students in tackling industry-specific problems. They contain interactive coding exercises with embedded hints to aid problem-solving and help students translate concepts into functional code. In this stage, students took on roles of developers, understanding the problem and creating their code, and deployers, refining and testing their solutions on the BBC micro:bits, which is a portable programmable device \cite{BBCmicrobit} and other available hardware, allowing students to see their code in action (see Figure \ref{fig:MICROBIT}). These roles were also rotated among the group members to ensure shared learning and collaboration.

\subsubsection{Collaborative Design Thinking Activities}
At this stage, groups have two hours to collaborate within their nation on a design thinking activity. The aim is to explore and reflect on the role of key technologies, i.e., AI, Robotics, and IoT, in their second assigned industry: tourism, food technology, and architecture. For example, Metal and Fire focus on solving challenges in the architecture industry. Water and Earth address issues within the food industry, while Wood tackles problems in tourism. Students are encouraged to take ownership of their industry-specific challenges to explore innovative solutions through applying these technologies (see Supplementary Material for a sample of the design thinking sheets). 

The design thinking activity has three parts. First, students are tasked with creating visual representations of how AI, Robotics, and IoT can contribute to addressing the identified challenges within their respective industries. After completing this part, nations visit each other to exchange feedback. These visits promote collaborative learning and encourage students to refine their ideas based on feedback.
\begin{figure}[h!]
\centering
\includegraphics[width=1\linewidth]{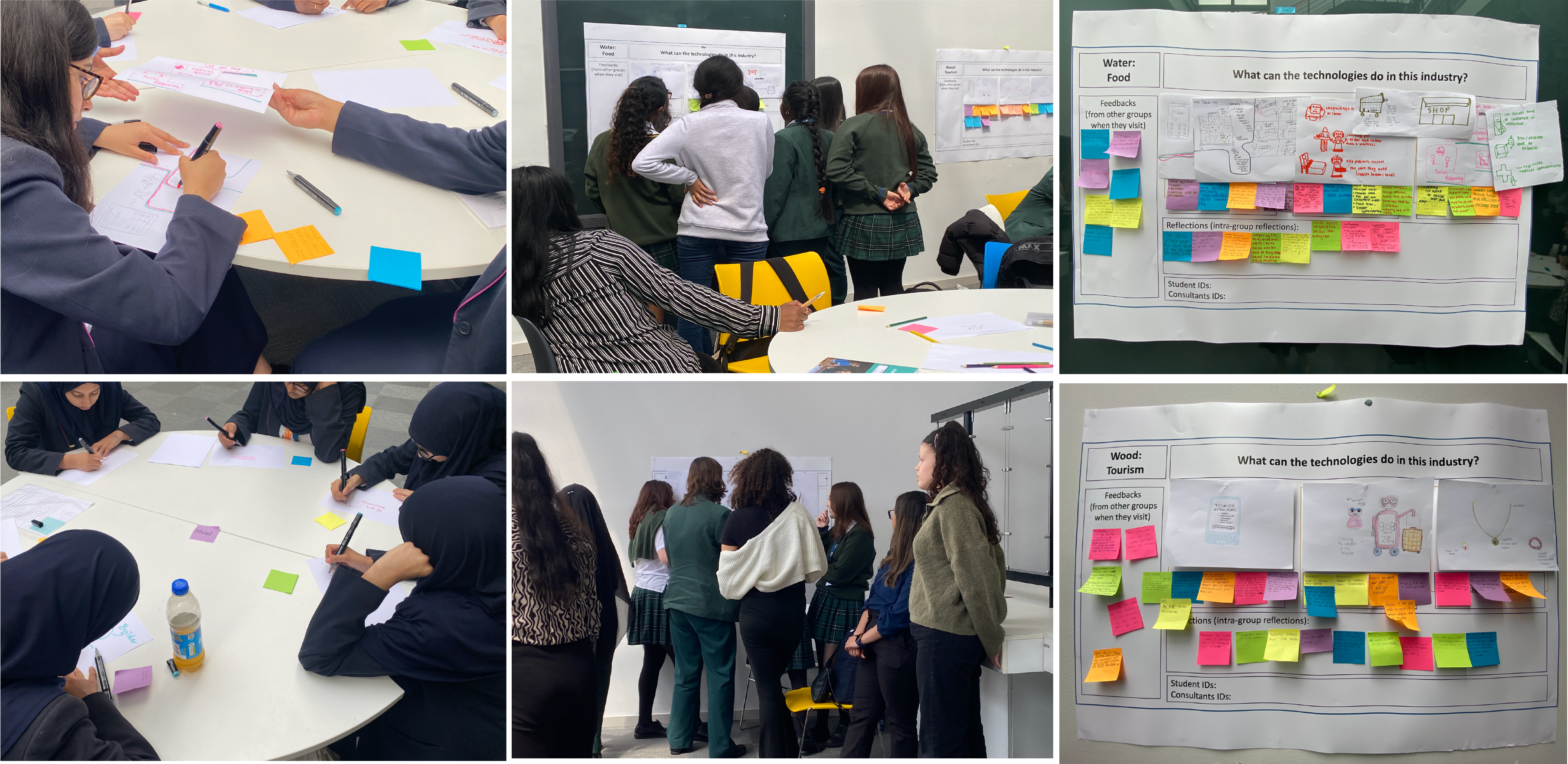}
\caption{Students use design thinking to explore various challenges linked to UN SDGs and different industries.}
\label{fig:DTA}
\end{figure}
The activity combines creative and analytical approaches, requiring students to visualise technological applications, articulate their ideas, and engage in collaborative reflection (see Figure \ref{fig:DTA}). Through this activity, the students connect abstract technological concepts to practical, real-world applications in the industry by focusing on AI, Robotics, and IoT.

\section{Research Design} 
\label{Research Design}
\begin{figure}[h!]
    \centering
    \includegraphics[width=1\linewidth]{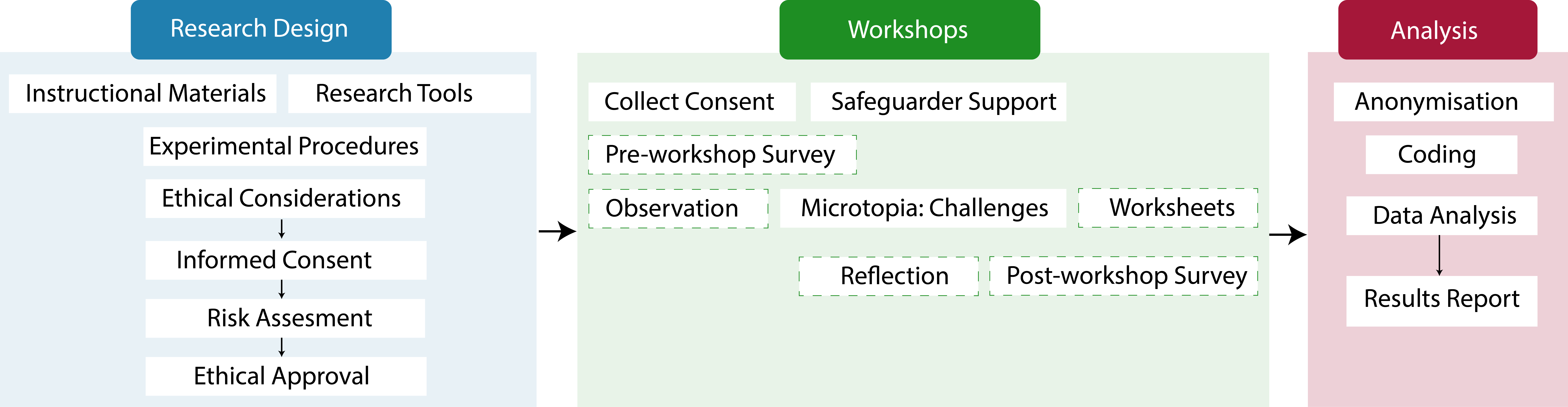}
    \caption{Methodological framework}
    \label{fig:CFS}
\end{figure}

With the factors and motivations mentioned earlier, this study aims to investigate the socioeconomic and ethnocultural factors that shape young girls’ preconceptions of Computer Science (CS), and evaluate the impact of the Microtopia programme as an intervention to shift these perceptions. The research framework contains research design, workshops (i.e., the Microtopia programme, delivered to groups of pupils), and analysis of the effectiveness of the programme (see Figure \ref{fig:CFS}). The research design covers the design of the instructional materials, the research tools such as questionnaires and worksheets, the detailed experimental procedures, along with considering ethical aspects and ensuring ethical compliance (which includes consent collection and safeguarding).  
To answer the research questions, data were collected via a pre-workshop questionnaire (see \ref{appendix:presurvey}), hands-on activity worksheets, and a post-workshop questionnaire (see \ref{appendix:postsurvey}) with reflections (see Figure \ref{fig:FRDQOR.png}). The pre-workshop questionnaire captures socioeconomic and ethnocultural factors, which may affect previous exposure to CS, and enables an understanding of pupils' beliefs around self-efficacy, perceptions, and attitudes towards CS.
\begin{figure}[h!]
    \centering
    \includegraphics[width=1\linewidth]{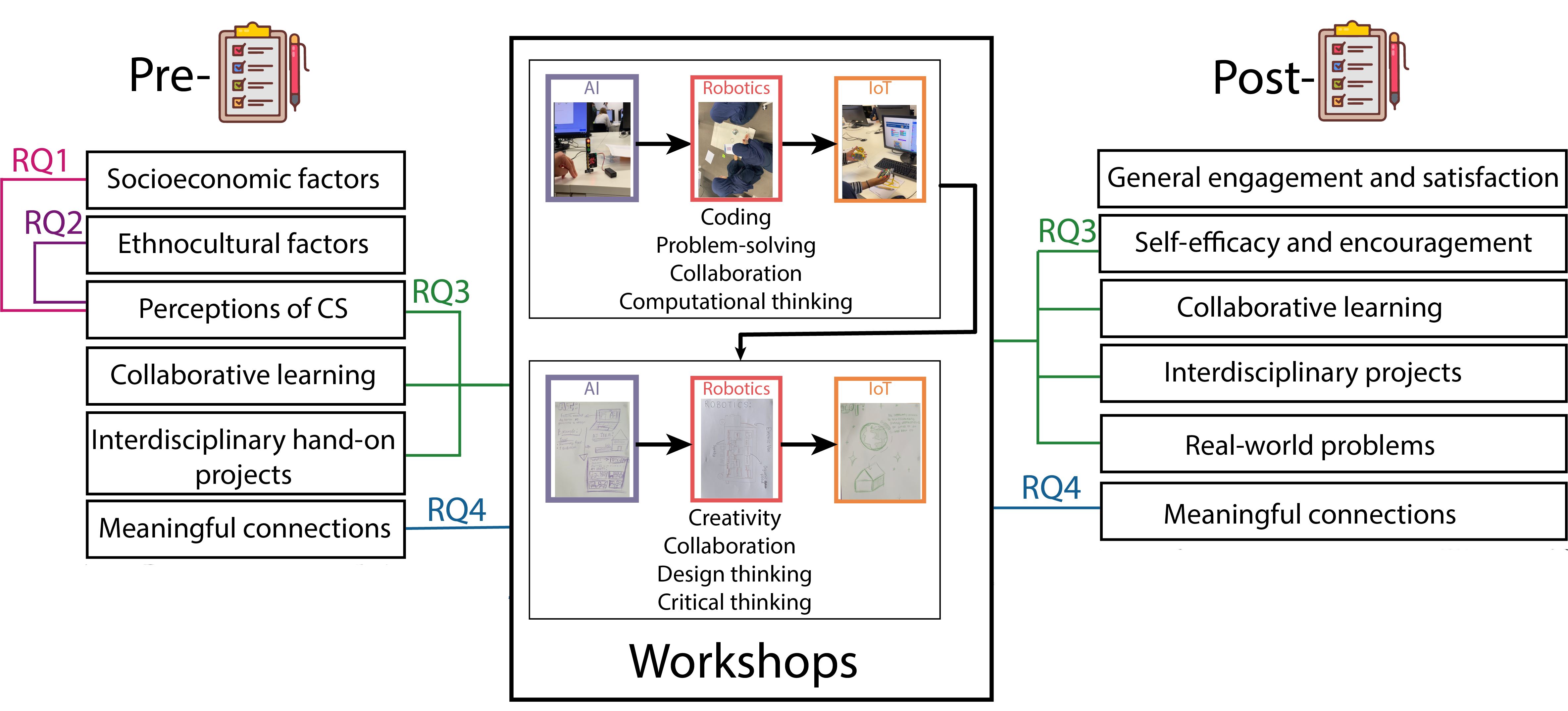}
    \caption{Mapping of RQ1–RQ4 to pre/post questionnaires and workshop design, with students completing coding tasks in AI, Robotics, and IoT, then using design thinking to apply the same technologies in a different industry.}
    \label{fig:FRDQOR.png}
\end{figure}

During the workshop, participants work on computing challenges and fill out worksheets designed to reinforce their learning. Facilitators observe participant interactions, behaviours, and engagement. The post-workshop data collection includes programme reflections and a followup questionnaire to measure shifts in perceptions and attitudes.
\subsection{Ethics and Workshop Procedures}
The study received ethical approval from the university and followed the guidelines in recruiting participants, consent and safeguarding, data collection and retention, in compliance with the UK's Data Protection Act 2018 \cite{DataProtectionAct2018}. Participants were recruited from secondary schools in a metropolitan area in the UK, an area known for its multicultural population and wide range of ethnic minority communities, which contribute significantly to the region’s character and diversity. A total of 90 pupils from three secondary schools participated voluntarily. The pupils, aged 12 to 15, were in Years 8 to 10 under the UK school system. Each workshop hosted 30 pupils for a one-day session. All collected data was anonymised.
Teachers from the participating schools confirmed participant eligibility and obtained parental consent before the study. During the workshop, pupils were briefed that they could withdraw from the study at any time. Their well-being was ensured in accordance with the university's safeguarding guidelines.

\subsection{Questionnaire Design}
Both the pre-workshop and post-workshop questionnaires include multiple choice questions, written-answer questions, and rating questions. All rating questions are on a 5-point Likert scale, with 1 representing the lowest rating and 5 the highest. This structure has been used on all scales to express complex concepts in the form of consistent, clear, and understandable statements. The phrasing was adapted to the children’s age, making use of examples from \cite{Tai2022} and the Children’s Computer Use Scale \cite{Frantom2002}.

\subsubsection{Pre-Workshop Questionnaire Development and Reliability}
The internal consistency of the questionnaire's items scale was measured using Cronbach's alpha, yielding $\alpha =0.86$, which suggests acceptable reliability. An additional open-ended question was included to capture qualitative insights into influences on pupils' interest in CS. At a high level, this questionnaire is structured into five components, which we outline below.

\textit{Part I. Socioeconomic factors.} This section includes fifteen items on socioeconomic status that were generated from several resources (e.g. \citet{ONS2018,hasebrink2019-social}) to address the complexity of the subject and its multifaceted nature, taking into account socioeconomic status, informal access to computers, perceptions and general encouragement to study CS.

\textit{Part II. Ethnocultural factors.} This section includes twenty-seven items that explores the relationship between ethnicity, social background, and exposure to computers, since these factors can influence pupils’ perceptions and engagement with technology. The standard UK Ethnicity Categories were used \cite{ONS2021Diversity} along with questions exploring family immigration history.

\textit{Part III. Perceptions of CS.} This section includes thirteen items, and was developed based on related literature on pupils’ attitudes and views concerning CS \citep{Hur2017,sharma2021-girls,varma2010-selfefficacy,zdawczyk2022-girls}. It addresses confidence, interest, and stereotypes about CS.

\textit{Part IV. Collaborative learning.} This section includes twelve items that were developed based on relevant literature on group work and collaboration in educational settings \citep{Vrieler2021-club, Scott2023-comparing} to reflect various dimensions of collaboration, including teamwork, enjoyment, inclusivity, and perceived group dynamics in gendered groups. 

\textit{Part V. Interdisciplinary hand-on projects.} We developed two questions that investigated whether pupils thought that CS is an interdisciplinary subject.

\textit{Part VI. Meaningful connections.} This section includes seven items that were developed to investigate perceptions of how CS would solve real-world problems \citep{bhatt15-media}.

\subsubsection{Post-Workshop Questionnaire Development and Reliability}
The post-workshop questionnaire was designed to align with the corresponding sections with the pre-workshop questionnaire. At a high level, this questionnaire is structured into five components which we outline below.

\textit{Part I. Self-efficacy and encouragement.} This section includes eleven items that were created to explore pupils’ confidence, interest, and perceived challenges related to CS after engaging with computing tasks.

\textit{Part II. Collaborative and interdisciplinary learning aspects.} This section includes thirteen items aimed to assess pupils’ feelings, preferences, and attitudes after working together in teams on computing-related activities regarding the gender effect. The design of these questions draws from research on teamwork and group dynamics, teamwork satisfaction, communication, and inclusivity within the group. 

\textit{Part III. Real-world problems.} This section includes nine items that sought to understand pupils’ perceptions of how CS relates to real-world applications when they had the chance to work on such activities. This component is designed to evaluate how pupils see the subject’s relevance to everyday issues and global challenges.

\textit{Part IV. Meaningful connections.} This section includes seven items that were developed to measure pupils’ emotional connection to CS, which reflect on meaningful learning experiences and the importance of purpose in education.

\subsubsection{Data Analysis}
These variables were categorised into binary or grouped classifications for analysis. Responses marked as \textit{"Prefer not to say"} were assigned a value of $0$, ensuring their inclusion in the dataset while maintaining analytical consistency. Descriptive statistics summarise participants’ questionnaire responses. The T-test was used to compare the means of two groups under the assumption of normality and equal variances, testing the null hypothesis that their means are identical. A $p<0.05$ was set to indicate a statistically significant difference for our hypothesis tests. For paired non-parametric data, the Wilcoxon signed-rank test \citep{Woolson2008} evaluates differences between matched samples. Only statistically significant findings ($p<0.05$) are reported, in order to emphasise meaningful results. When normality is violated, the Kruskal--Wallis (H) \citep{Kruskal01121952} test assesses differences in medians across multiple groups, with significance determined by $p<0.05$ in SPSS. For two independent non-parametric groups, the Mann--Whitney U-test \citep{Mann1947OnAT} evaluates rank differences between two groups, where a $p<0.05$ suggests a significant distinction. Also for those tests, only statistically significant findings ($p<0.05$) are reported, to emphasise meaningful results. Regression analysis was performed in SPSS when the data is ordinal \citep{osborne2014best,osborne2016regression,field2024discovering}.

\section{Results}
\label{Results}
\noindent In the following four subsections, we will discuss our results with respect to each of the four primary research questions (RQ1-RQ4), respectively.
\subsection{Socioeconomic Factors}
This section addresses RQ1: \textit{To what extent do socioeconomic factors influence ethnic minority girls’ preconceptions of CS?}

\begin{figure}[h!]
\centering
\includegraphics[width=1\linewidth]{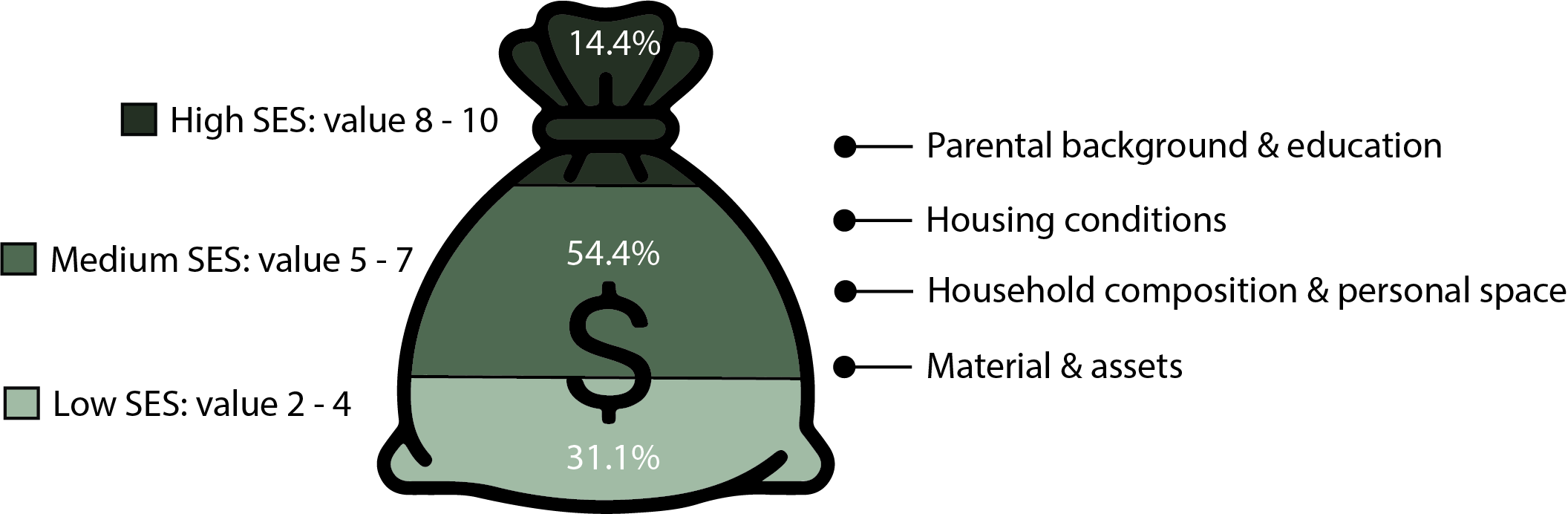}
\caption{Overview of composite socioeconomic status (SES) index and classification used in our study. Scores were summed and categorised as Low (2–4; 31.1\% of the participant population), Medium (5–7; 54.4\%), and High (8–10; 14.4\%), with darker shading indicating higher SES.}
\label{fig:SESvalue}
\end{figure}

Data were collected on parental employment, parental education, household bedrooms, siblings, home ownership, personal bedroom access, and car ownership. Variables were coded as follows: parental higher education (Yes = 1, No = 0), bedrooms (1 = 0, 2 = 1, 3 = 2, 4 = 3, 5+ = 4), siblings (0 = 2, 1 = 1, 2 = 0, 3+ = –1), home ownership (Own = 2, Rent = 0), personal bedroom (Yes = 1, No = 0), and car ownership (Yes = 1, No = 0). A composite measure of socioeconomic status (SES) was employed to minimise potential bias associated with using a single indicator. The composite SES score was structured such that a higher value indicated a higher socioeconomic status. The classification criteria were adapted from \cite{oakes2017measurement} and pupils were categorised into three SES groups (see Figure \ref{fig:SESvalue}), where a score of 2--4 was classified as low ($N = 28$), a score of 5--7 was classified as medium (N = 49), and a score of 8--10 was classified as high ($N = 13$).I don't understand how variability in SES scores was increased.

Given that the assumption of normality was violated and the SES groups were of unequal size, a Kruskal--Wallis H test was conducted to examine differences in the SES composite score across three groups (low, medium, high). The Kruskal--Wallis H test showed significant differences across several questionnaire items. For the \textit{CS Preconceptions} section: Significant group differences were observed for the statement “I don’t mind being the only girl in a group” ($H(2) = 11.50, \; p = 0.003$), as well as for “I do not think I can work well if I am the only girl in a group” ($H(2) = 10.54, \; p = 0.005$). Further, responses to “I think that computer science is a boys’ club” also differed significantly by SES group ($H(2) = 8.48, \; p = 0.014$). Finally, pupils’ perceptions that “computer science is related to what I experience in the real world applications” showed significant differences ($H(2) = 7.36, \; p = 0.025$). For the \textit{Collaboration \& Teamwork} section: Pupils also reported that they were okay with doing different jobs in their group when needed ($H(2) = 7.81, \; p = 0.020$), and that boys and girls can work together well in a group ($H(2) = 8.62, \; p = 0.013$). Table \ref{tab:Kruskal} shows the significant differences that emerged between SES groups on six questionnaire items, along with the descriptive group means.
\begin{table}[h!]
\centering
\caption{Mean (SD) responses to the six survey items across SES groups (Low, Mid, High)}
\begin{tabular}{p{6cm}ccc}
\hline
\multirow{2}{*}{Item} & \multicolumn{3}{c}{Mean (SD)} \\
 & Low (N=28) & Medium (N=49) & High (N=13) \\
\hline
\multicolumn{4}{l}{\textbf{CS Preconceptions}} \\
\small{Cannot work well if I'm the only girl} & 3.2 (1.6) & 2.1 (1.16) & 2.1 (1.26) \\
\small{CS is a boys’ club} & 1.8 (1.17) & 1.2 (0.43) & 1.2 (0.38) \\
\small{CS relates to real world} & 3.5 (0.96) & 3.0 (1.17) & 4.0 (1.08) \\
\small{Don’t mind being the only girl} & 2.8 (1.47) & 3.4 (1.35) & 4.4 (0.96) \\
\hline
\multicolumn{4}{l}{\textbf{Collaboration \& Teamwork}} \\
\small{Okay with doing different jobs within the group} & 3.8 (0.88) & 4.3 (0.73) & 4.5 (0.52) \\
\small{Boys and girls can work together} & 3.8 (1.09) & 4.5 (0.68) & 4.5 (0.78) \\
\hline
\end{tabular}
\label{tab:Kruskal}
\end{table}
Beyond the composite SES score, we also investigated whether individual SES indicators predicted preconceptions, using Mann--Whitney U tests. We found consistent patterns across all six preconception questions and across the three SES group comparisons. For the \textit{CS Preconceptions} section, pupils in the low SES group were more likely to agree that CS is a boys' club than those in the medium SES group ($U = 494$, $p = 0.006$). Low SES pupils were also less confident that they could work well if they were the only girl in the group compared with both the medium SES group ($U = 403$, $p = 0.002$) and the high SES group ($U = 105$, $p = 0.03$). On the other hand, pupils in the high SES group reported being more comfortable with being the only girl in a group than both the low SES group ($U = 67$, $p < 0.001$) and the medium SES group ($U = 163$, $p = 0.005$). In addition, high SES pupils felt that CS was more related to their real-world experiences compared with the medium SES group ($U = 179.50$, $p = 0.013$).  

\begin{figure}[h!]
\centering
\includegraphics[width=1\linewidth]{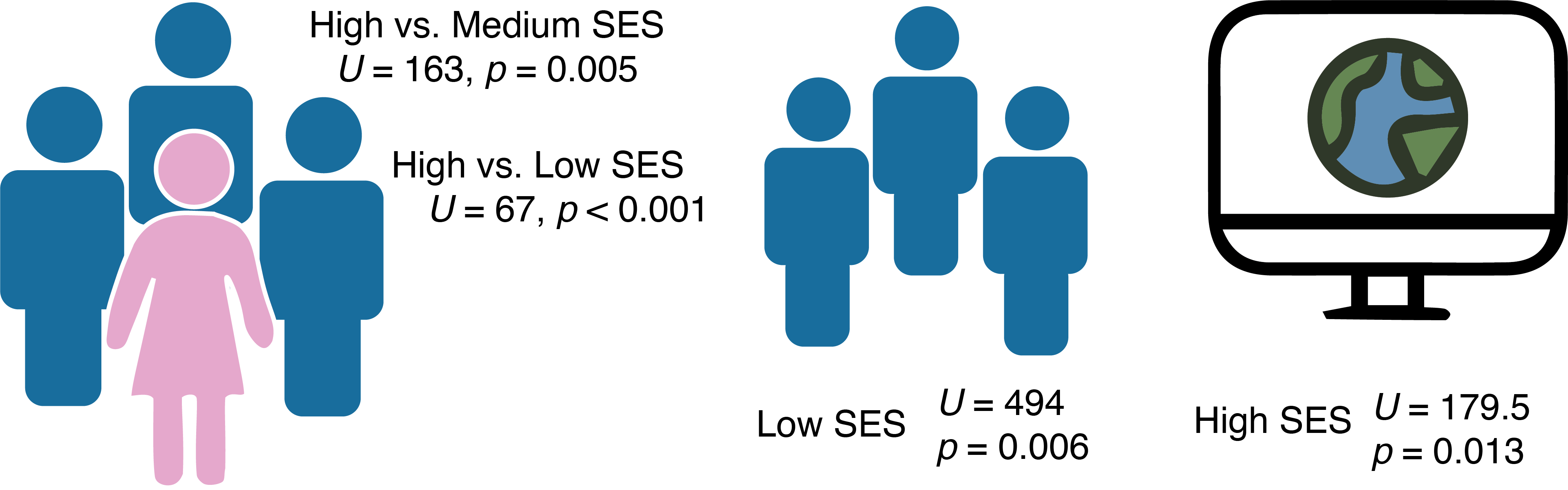}
\caption{\textit{Left:} High SES pupils were more comfortable being the only girl in a group, while low SES pupils were less comfortable. \textit{Middle:} Low SES pupils were more likely to see computer science as a boys’ club. \textit{Right:} High SES pupils were more likely to see computing as connected to real-world experiences.}
\label{fig:Mann-WhitneyResults}
\end{figure}

For the \textit{Collaboration \& Teamwork} section, pupils in the medium SES group ($U = 479$, $p = 0.02$) and the high SES group ($U = 102$, $p = 0.02$) were more likely than those in the low SES group to agree that doing different jobs within a group is important. Moreover, pupils in the medium SES group were more likely than those in the low SES group to think that girls and boys work well together in a group ($U = 440.50$, $p = 0.005$). These findings are illustrated in Figure ~\ref{fig:Mann-WhitneyResults}.

Moreover, additional Mann--Whitney U tests revealed several significant associations between individual SES indicators and pupils' perceptions (see Table \ref{tab.SESResultsComponents}).
\begin{table}[h!]
\centering
\caption{Associations between SES components and pupils' responses to computer science perception questions. Comparisons were made between its two categories: home ownership: own vs.\ rent; parental higher education: yes vs.\ no; parental employment: employed vs.\ not employed; own bedroom: yes vs.\ no. The two rightmost columns show the Mann--Whitney U test statistic $U$ and corresponding $p$ values.}
\begin{tabular}{llcc}
\hline
SES Indicators & Item & $U$ & $p$ \\ \hline
\multicolumn{4}{l}{\textbf{CS Preconceptions}} \\ 
Home owner & Would not mind being the only girl  & $695$   & $0.008$ \\ 
Home owner & CS relates to real world applications & $761$   & $0.039$ \\
Parent has higher education & Would not mind being the only girl & $70$   & $0.033$ \\
Mom employed & Disagree with CS club is a boy's club & $363$ & $0.002$ \\
\hline
\multicolumn{4}{l}{\textbf{Collaboration \& Teamwork}} \\ 
Having own bedroom & Okay with doing different jobs in the group & $681$   & $0.004$ \\ 
Dad employed & Okay with doing different jobs in the group & $349$ & $0.013$ \\
\hline
\end{tabular}
\label{tab.SESResultsComponents}
\end{table}

\subsection{Ethnocultural Factors}
This section addresses RQ2: \textit{To what extent do ethnocultural factors influence ethnic minority girls’ preconceptions of CS?}

\begin{figure}[h!]
\centering
\includegraphics[width=0.9\textwidth]{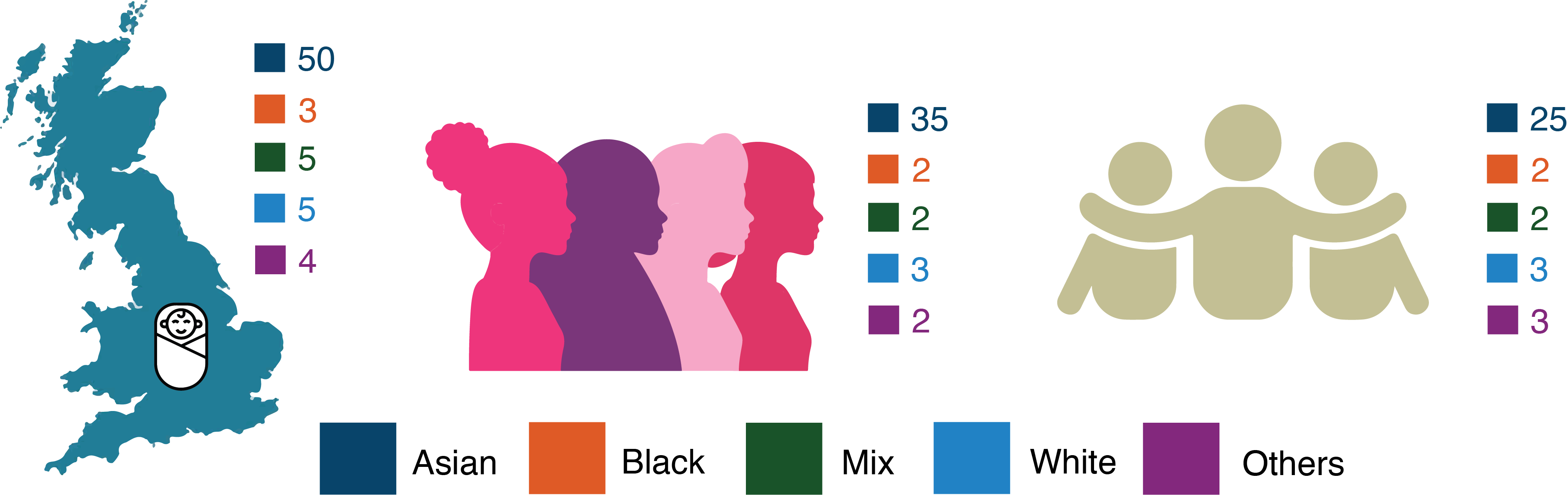}
\caption{Identity strength and community closeness by ethnicity. \textit{Left:} shows the number of participants from each ethnic group born in Great Britain. \textit{Middle} shows the number of participants reporting high identity strength, per ethnic group. 
\textit{Right:} shows the number of participants reporting high community closeness, per ethnic group.}
\label{fig.EthnicityIdentity}
\end{figure}

Data were collected on participants’ ethnicity and their CS preconceptions, collaboration, ease of communication, and shared efforts among teammates, as well as their perceptions of gender exclusivity in computer science clubs. The analyses were then conducted to examine the differences between ethnic groups. The ethnic groups were Asian or British Asian (shortened to Asian; $N=68$), Black ($N=6$), Mixed ($N=7$), White ($N=5$), and Others ($N=4$; comprising Arab and Traveller participants). See Figure \ref{fig.EthnicityIdentity}, which illustrates differences in identity strength and community closeness across ethnic groups in our sample. 

We conducted Kruskal--Wallis tests, which reveal significant overall differences between ethnic groups. We report the significant findings from a series of questionnaire items that were grouped into two categories: \textit{Computer Science Preconception} and \textit{Collaboration \& Teamwork}. The descriptive statistics (see Table~\ref{tab.ethnodescriptive}) summarise pupils’ preconceptions across ethnocultural groups. Mean scores were broadly similar, ranging from 2.81 (Mixed) to 3.36 (Other), with an overall mean of 3.28 (Standard Deviation $(\mathrm{SD}) = 1.58$, Standard Error $(\mathrm{SE}) = 0.06$). Although differences between groups were small, standard deviations indicated variability within groups. Standard errors highlighted the effect of sample size on the precision of estimates: the Asian group, representing the largest proportion of pupils, showed the most stable mean ($\mathrm{SE} = 0.07$), whereas smaller groups such as White $\mathrm{SE} = 0.24$) and Others ($\mathrm{SE} = 0.28$) come with less precise estimates.

Follow-up Mann-Whitney U tests have been conducted, which showed that Asian pupils differed significantly from Mixed pupils, with the strongest effect for ``I like it when everyone helps to finish work'' ($U = 81$, $p = 0.001$), which remained significant. Other Asian group vs. Mixed group differences emerged at $p < 0.05$, including their views on both boys and girls being good at computer science ($U = 131.5$, $p = 0.009$), feeling good about teamwork ($U = 98.5$, $p = 0.008$), ease of talking to teammates ($U = 98.5$, $p = 0.007$), comfort with group roles ($U = 105.5$, $p = .01$), whether CS is a boys’ club ($U = 139$, $p = 0.01$), and not working well in a group ($U = 123$, $p = 0.02$). Additional contrasts were observed in attitudes toward teamwork and collaboration. In particular, between the Asian vs. Black groups ($U = 92$, $p = 0.02$; $U = 76$, $p = 0.005$), Black vs. White groups ($U = 2.5$, $p = 0.017$), Black vs. Other groups ($U = 0$, $p = 0.01$), White vs. Mixed groups (both $U = 4.5$, $p = 0.03$), and Mixed vs. Other groups ($U = 1.5$--$3$, $p = 0.01$--$0.042$). We also investigate the generational status of pupils and their preconceptions of CS, please see \ref{appendix:GenerationalStatus}.
\begin{table}[h!]
\centering
\caption{Descriptive statistics (Mean, Standard Deviation (SD), Standard Error (SE)) for Computer Science Preconception and Collaboration \& Teamwork items, by ethnicity.}
\resizebox{\textwidth}{!}{%
\begin{tabular}{p{4.2cm}ccccc}
\hline
\textbf{Item} & \textbf{Asian} (75.6\%) & \textbf{Black} (6.7\%) & \textbf{Mixed}(7.8\%) & \textbf{White}(5.6\%) & \textbf{Other}(4.4\%) \\
\hline
\multicolumn{6}{l}{\textbf{CS Preconception}} \\
I think CS is for boys only & 1.13 (0.46, 0.06) & 1.67 (0.82, 0.34) & 1.43 (0.79, 0.30) & 1.00 (0.00, 0.00) & 1.25 (0.50, 0.25) \\
I think boys and girls can be good at CS & 4.78 (0.45, 0.05) & 5.00 (0.00, 0.00) & 3.71 (1.50, 0.57) & 4.80 (0.45, 0.20) & 5.00 (0.00, 0.00) \\
\hline
\multicolumn{6}{l}{\textbf{Collaboration \& Teamwork}} \\
I feel good about working with teammates & 4.03 (0.63, 0.08) & 3.33 (0.82, 0.34) & 2.86 (1.07, 0.40) & 4.40 (0.55, 0.25) & 4.25 (0.50, 0.25) \\
I think it is important to work together & 4.38 (0.65, 0.08) & 3.67 (0.52, 0.21) & 3.43 (0.79, 0.30) & 4.00 (0.71, 0.32) & 4.25 (0.50, 0.25) \\
I like when we all help each other to finish work & 4.46 (0.62, 0.08) & 3.50 (0.55, 0.22) & 3.00 (0.82, 0.31) & 4.20 (0.45, 0.20) & 4.25 (0.50, 0.25) \\
It is easy to share ideas with teammates & 4.38 (0.65, 0.08) & 3.50 (0.55, 0.22) & 2.86 (1.07, 0.40) & 4.40 (0.55, 0.25) & 4.50 (0.58, 0.29) \\
I am okay with doing different jobs in groups & 4.18 (0.67, 0.09) & 4.17 (0.41, 0.17) & 3.29 (1.11, 0.42) & 4.40 (0.55, 0.25) & 4.75 (0.50, 0.25) \\
I think CS is a boys’ club & 1.32 (0.61, 0.08) & 1.67 (0.82, 0.34) & 2.00 (0.82, 0.31) & 1.20 (0.45, 0.20) & 1.25 (0.50, 0.25) \\
I do not think I work well in a group & 1.72 (0.79, 0.10) & 2.17 (0.98, 0.40) & 2.57 (1.13, 0.43) & 1.20 (0.45, 0.20) & 1.00 (0.00, 0.00) \\
\hline
\end{tabular}%
}
\label{tab.ethnodescriptive}
\end{table}

\subsection{Perceptions of CS}
This section addresses RQ3: \textit{To what extent does Microtopia influence ethnic minority girls’ perceptions of Computer Science (CS)?}

We examine ratings in several areas: \textit{Perceptions of CS}, attitudes toward \textit{Collaboration \& teamwork}, the \textit{Impact of the Interdisciplinary projects}, and whether pupils envisioned CS as part of their \textit{Future Career}. Pre- and post-workshop questionnaire responses were compared, and the descriptive statistics in Table \ref{tab:CSWilcoxonSignedTestAll} summarise pupils’ views before and after the workshop, where $N = 90$. The data were treated as ordinal, and a Wilcoxon Signed Rank Test was conducted. The results indicated significant positive shifts in pupils’ confidence, enjoyment, interest, and perceived ability in computer science ($Z = –6.22$ to $–7.44$, $p < 0.05$). Pupils also reported more comfort in sharing ideas and collaborating with teammates ($Z = –7.04$ to $–7.31$, $p < 0.001$), as well as flexibility in taking on different roles ($Z = –6.64$, $p = 0.036$). In terms of interdisciplinary and wider impact, pupils showed increased recognition that computer science requires in multiple fields of study ($Z = –6.23, p = 0.028$) and that CS can help address global challenges ($Z = –6.86, p = 0.001$). However, no significant changes were observed in intention to pursue computer science at university ($Z = –5.18, p = 0.080$) or as a career ($Z = –6.67, p = 1.000$), nor in views on the importance of teamwork ($Z = –6.95, p = 0.194$), or applicability of computer science to everyday life ($Z = –5.78, p = 0.249$). 
\begin{table}[h!]
\centering
\small
\caption{Pre- and Post-Microtopia workshop results with Wilcoxon Signed Rank Test (Mean (M), SD, SE, Z, p).}
\label{tab:CSWilcoxonSignedTestAll}
\resizebox{\textwidth}{!}{%
\begin{tabular}{p{7cm}ccccc}
\hline
\textbf{Item} & \textbf{Pre-test M (SD, SE)} & \textbf{Post-test M (SD, SE)} & \textbf{Z} & \textbf{p} \\
\hline
\multicolumn{5}{l}{\textbf{CS Perceptions}} \\
I feel confident in my skills when using a computer. & $3.47 \,(1.02,\,0.11)$ & $4.23 \,(0.84,\,0.09)$ & $-7.39$ & \textbf{$p < 0.001$} \\
I believe computer science is enjoyable. & $3.18 \,(1.17,\,0.12)$ & $3.99 \,(0.98,\,0.10)$ & $-7.39$ & \textbf{$p < 0.001$} \\
I am interested in studying computer science. & $3.09 \,(1.24,\,0.13)$ & $3.82 \,(1.07,\,0.11)$ & $-7.17$ & \textbf{$p < 0.001$} \\
I feel good about learning/studying computer science. & $3.44 \,(1.10,\,0.12)$ & $3.78 \,(0.97,\,0.10)$ & $-6.22$ & \textbf{$p = 0.010$} \\
I am confident in my ability to do well in computer science courses. & $3.32 \,(1.05,\,0.11)$ & $3.81 \,(1.04,\,0.11)$ & $-7.44$ & \textbf{$p < 0.001$} \\
I want to have a job someday that involves computers. & $3.20 \,(1.17,\,0.12)$ & $3.39 \,(1.22,\,0.13)$ & $-6.56$ & \textbf{$p = 0.038$} \\
I believe anyone can learn computer science. & $4.52 \,(0.72,\,0.08)$ & $4.30 \,(0.92,\,0.10)$ & $-7.39$ & \textbf{$p = 0.042$} \\
I believe that learning computer science is hard. & $3.42 \,(1.02,\,0.11)$ & $2.74 \,(1.14,\,0.12)$ & $-6.75$ & \textbf{$p < 0.001$} \\
People who do computer science need to be very smart. & $3.30 \,(1.10,\,0.12)$ & $3.01 \,(1.14,\,0.12)$ & $-6.07$ & \textbf{$p = 0.047$} \\
\hline
\multicolumn{5}{l}{\textbf{Collaboration \& Teamwork}} \\
I feel good about working with my teammates on projects. & $3.92 \,(0.99,\,0.10)$ & $4.33 \,(0.95,\,0.10)$ & $-7.04$ & \textbf{$p < 0.001$} \\
It’s easy for me to talk to my teammates and share my ideas when we’re working on something. & $3.96 \,(1.07,\,0.11)$ & $4.36 \,(0.93,\,0.10)$ & $-7.31$ & \textbf{$p < 0.001$} \\
I am okay with doing different jobs in our group when we need to. & $4.14 \,(0.79,\,0.08)$ & $4.34 \,(0.81,\,0.09)$ & $-6.64$ & \textbf{$p = 0.036$} \\
I think boys and girls can work together well in a group. & $4.28 \,(0.89,\,0.09)$ & $3.42 \,(1.53,\,0.16)$ & $-7.43$ & \textbf{$p < 0.001$} \\
I would not mind being the only girl in a group. & $3.46 \,(1.81,\,0.19)$ & $2.51 \,(1.44,\,0.15)$ & $-5.40$ & \textbf{$p = 0.001$} \\
I do not think I can work well if I am the only girl in a group. & $2.42 \,(1.41,\,0.15)$ & $1.27 \,(0.85,\,0.09)$ & $-7.28$ & \textbf{$p < 0.001$} \\
\hline
\multicolumn{5}{l}{\textbf{Interdisciplinary}} \\
I think that computer science requires different subjects and fields, like science, art, physics, design, social sciences, or maths. & $4.00 \,(0.91,\,0.10)$ & $4.24 \,(0.94,\,0.10)$ & $-6.23$ & \textbf{$p = 0.028$} \\
\hline
\multicolumn{5}{l}{\textbf{Future in CS}} \\
I want to learn more about computer science in the future & $3.99 \,(0.93,\,0.10)$ & $4.29 \,(0.80,\,0.08)$ & $-6.45$ & \textbf{$p = 0.011$} \\
I think I can be successful in computer science if I work hard & $3.73 \,(1.01,\,0.11)$ & $4.21 \,(0.85,\,0.09)$ & $-7.45$ & \textbf{$p < 0.001$} \\
\hline
\end{tabular}
}
\end{table}
Asian pupils showed the broadest pattern of improvements across all sections. Confidence in computing rose from a mean of $3.41$ to $4.37$ ($p<0.001$), and interest in studying computer science rose from $3.04$ to $4.50$ ($p<0.001$). Negative perceptions declined sharply: the belief that ``CS is hard'' dropped from a mean of 3.50 to 1.65 ($p<0.001$), and that ``CS is only for very smart people'' from $3.35$ to $1.84$ ($p<0.001$). Collaboration also strengthened, with reduced discomfort about being the only girl in a group ($2.43$ to $1.22$, $p<0.001$) and a marked shift away from preferring to work alone ($2.38$ to $4.29$, $p<0.001$). Pupils more strongly recognised CS as interdisciplinary ($4.06$ to $4.32$, $p=0.049$), expressed greater aspirations to keep learning ($4.09$ to $4.40$, $p=0.022$), and felt more confident in their ability to succeed ($3.74$ to $4.32$, $p<0.001$). Overall, this group exhibited the most consistent and wide-ranging positive changes.

\begin{figure}[h!]
\centering
\includegraphics[width=1\textwidth]{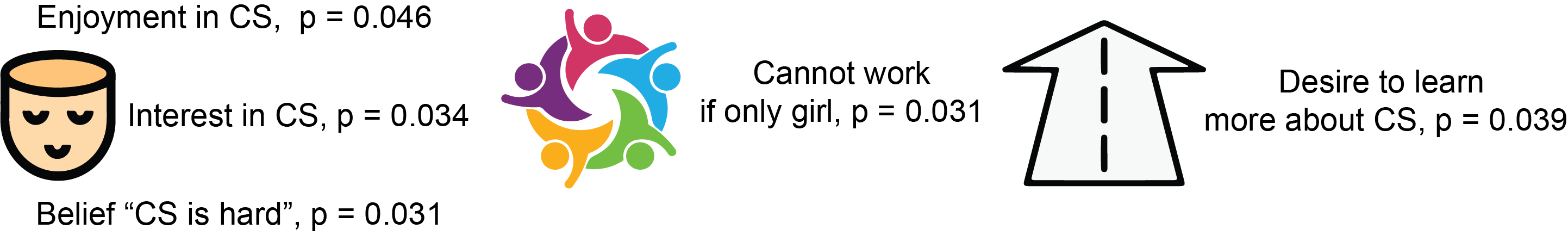}
\caption{After the Microtopia workshop, Black pupils showed meaningful improvements in their perceptions, including increased enjoyment, reduced belief that ``CS is hard'', higher confidence in teamwork, and a stronger desire to learn more about CS.}
\label{fig.BlackPerceptionsM}
\end{figure}

Black pupils also demonstrated positive shifts, though across a narrower set of areas. Interest in studying CS increased from a mean of $3.33$ to $4.33$, and enjoyment of learning CS rose from $3.33$ to $4.00$. Negative perceptions decreased, with the belief that ``CS is hard'' falling from $3.00$ to $1.50$. In collaboration, pupils expressed less concern about being the only girl in a group, dropping from a mean of $3.17$ to $1.00$, indicating confidence in inclusive teamwork (see Figure \ref{fig.BlackPerceptionsM}). 

For pupils in the \emph{Other} group, changes were non-significant, reflecting the very small group size, although descriptive statistics indicated trends consistent with the overall cohort.

\subsection{Meaningful Connections and Engagement in Computing}  
This section addresses RQ4: \textit{To what extent does Microtopia enable ethnic minority girls to establish meaningful connections between CS and real-world challenges?}

Data on meaningful connections were collected and analysed in relation to self-reported engagement. Since the Likert-scale responses were ordinal, Kendall’s Tau correlations were first computed (see Table \ref{tab:correlation-categories}), followed by ordinal regression analysis to examine predictive effects. The correlations indicate consistent positive associations. Pupils who cared about the problems embedded in the tasks reported stronger motivation to solve them ($\tau = 0.40, p < 0.001$). This suggests that pupils who cared about solving problems were more likely to believe in the positive role of computers in addressing global challenges. Pupils also felt that working on CS projects that were aimed at improving the world would enhance their happiness and motivation. There was also a belief that computers can solve issues like pollution was positively correlated. The relevance of the workshop concept based on the UNSDG was also strongly correlated. There was also an intent to pursue a job that uses computer skills for environmental benefit. 
\begin{table}[h!]
\centering
\caption{Significant Kendall’s Tau correlations between engagement of the \textit{Microtopia} workshop and perceptions of meaningful connections in computing projects.}
\begin{tabular}{p{9cm} c c}
\hline
\textbf{Statement} & \textbf{Kendall's $\tau$} & \textbf{p-value} \\ \hline

\multicolumn{3}{l}{\textbf{Motivation and Engagement}} \\
I cared about the problem in the tasks, so I wanted to solve it more. & $0.40$ & $p < 0.001$ \\
Knowing I’m helping the world with my computer skills would make me try even harder on my tasks. & $0.35$ & $p < 0.001$ \\
\hline
\multicolumn{3}{l}{\textbf{Belief in Computing’s Social Impact}} \\
I think computers can help make the world a better place by solving problems like pollution. & $0.25$ & $p = 0.007$ \\
Working on a computer project that helps make the world better would make me really happy. & $0.35$ & $p < 0.001$ \\
Working on a project that helps the world using computers makes me feel really powerful. & $0.28$ & $p = 0.001$ \\
\hline
\multicolumn{3}{l}{\textbf{Relevance to Global Issues and Future Pathways}} \\
The workshop idea built on United Nations Sustainable Development Goals was meaningful and relevant. & $0.34$ & $p < 0.001$ \\
After today’s workshop, I feel like I might want to have a job where I use computers to help the Earth. & $0.30$ & $p = 0.001$ \\
\hline
\end{tabular}
\label{tab:correlation-categories}
\end{table}

\textit{Ordinal Regression Analysis:} Assumptions were tested and indicated there was no multi-collinearity ($VIF < 5$) and the test of parallel lines produced a non-significant value, so the assumption of proportional odds has been met. Based on the model fitting data, we can say that there was a significant improvement in model fit over the null model $H(7) = 24.79, p < 0.001$, reflecting the seven predictors from the questionnaire items. The Goodness of fit model showed non-significant results, showing that the model fits the data well \citep{petrucci2009primer} where Pearson chi-square test $H(171.934) = 171.93$, $p = 0.063$ and so does the deviance test $H(134.236) = 136.236$, $p = 0.687$. These results suggest good model fit. These findings suggest that the model represents the relationships between the data, providing support for the importance of using computer skills to address societal challenges.

\section{Discussion}
\label{Discussion}
\subsection{Implications}
The findings from this study highlight the potential of interdisciplinary, real-world problem-solving approaches in fostering interest and engagement among ethnic minority female pupils in Computer Science (CS). The \textit{Microtopia} programme, which integrates AI, IoT, and Robotics with the UN SDGs, provides an engaging framework that links computing with global challenges, potentially increasing pupils' motivation to pursue CS-related fields. 

This study goes beyond introducing a learning programme to examine the socioeconomic and ethnocultural factors that shape pupils’ perceptions of computer science. Our results suggest that higher SES pupils tend to express confidence in CS, while lower SES pupils are more likely to perceive it as a male-dominated discipline. This underscores the need for targeted interventions that address disparities in confidence, exposure, and perception of CS among pupils from different socioeconomic backgrounds. We also examined immigration generational status and found that longer family settlement in the UK was positively associated with pupils’ perceptions of computer science, teamwork, and collaboration.

Despite the workshop’s positive effects on pupils’ perceptions of computer science, statistical analysis revealed no significant pre- to post-workshop differences in pupils’ intentions to pursue a career in CS. This indicates that a single short-term intervention may not be sufficient to fundamentally shift attitudes toward CS. This suggests that longer-term, sustained engagement is required to foster meaningful changes in perception. Moreover, this study highlights that pupils showed higher engagement when CS concepts were linked to meaningful applications, reinforcing the importance of purpose-driven learning experiences. Thus, our findings support the integration of interdisciplinary, socially relevant computing projects and role-playing activities into computer science curricula to expose prospective pupils to CS in an effective way. 

\subsection{Limitations}
While this study provides valuable insights, several limitations should be acknowledged. First, the sample of 90 pupils, although diverse, may not be fully representative of all ethnic minority female pupils in the broader educational landscape. The study primarily focuses on pupils in a metropolitan area in the UK, and the findings may not be generalisable to different cultural or regional contexts. Future studies could expand the sample size and include participants from multiple geographic locations to potentially confirm the external validity of our conclusions. Second, this study relies on self-reported survey responses, which are subject to social desirability bias. pupils may have provided responses that reflect socially acceptable attitudes rather than their genuine perceptions of CS. Future research could incorporate qualitative methods, such as interviews or focus groups, to gain deeper insights into pupils' personal experiences and motivations in CS education.

Finally, this study primarily assessed pupils' perceptions and confidence levels, and did not include actual skill development or learning outcomes in CS in its scope. Future research could incorporate performance-based assessments to evaluate the impact of such interdisciplinary workshops on pupils' computational thinking and problem-solving abilities.

\section{Conclusion}
\label{Conclusion}
This paper studies the persistent underrepresentation of ethnic minority women in computer science and the systemic barriers they face, including stereotypes, lack of early exposure, and limited role models. Through an interdisciplinary and collaborative approach, the \textit{Microtopia} programme successfully engaged pupils by integrating computing with real-world problems. The findings show that pupils from higher socioeconomic backgrounds display greater confidence in CS, while lower SES pupils were more likely to view CS as a male-dominated field. Additionally, our results show that ethnocultural differences influence pupils’ perceptions of teamwork, collaboration, and their sense of belonging in CS.

One of the key contributions of this study is the demonstration that interdisciplinary, collaborative, problem-driven learning can positively impact pupils’ engagement and attitudes toward CS. The \textit{Microtopia} workshops significantly improved participants’ confidence, interest, and enjoyment in computing, reinforcing the importance of purpose driven education. The study also found that pupils' motivation improved when they saw CS as a tool for addressing sustainability and global challenges. This suggests that shifting CS education toward meaningful, impact-driven learning could be an effective strategy for increasing diversity and inclusion in the field.

By providing empirical evidence on how socioeconomic and ethnocultural factors shape CS preconceptions, this paper contributes valuable insights into designing more inclusive education for computer science. The study demonstrates the need for targeted interventions that foster a sense of belonging, increase early exposure, and connect CS to broader societal goals. Future work and follow-up work on this topic should explore long-term impacts of such initiatives, and how this approach can be scaled to further empower ethnic minority female pupils with an enriched perspective in STEM. We furthermore plan to analyse and reflect on qualitative aspects of the \textit{Microtopia} project (rather than the quantitative aspects, as covered in the present paper) as follow-up work.
\bibliographystyle{elsarticle-harv} 
\bibliography{GenderResearch}
\appendix
\section{Pre-workshop Questionnaire}
\label{appendix:presurvey}
\paragraph{Socioeconomic Status}
\begin{itemize} 
\item Are your parents employed?
\item	What does your mother do?
\item What does your father do?
\item Do any of your parents have higher education qualifications?
\item	How many bedrooms do you have in your home?
\item	How many siblings do you have?
\item	Do your family own the house that you are living in?
\item	Do you have your own bedroom for yourself?
\item	Does your family have a car, van or truck?
\item	Does your family own a computer?
\end{itemize}

\paragraph{Ethnic Background, Generational Influence, and Exposure to Technology}
\begin{itemize}
 \item What is your ethnicity?
 \item	Were you born in the UK?
\item	Which generation of your family immigrated to the UK?
\item	How strongly do you identify with your ethnic group?
\item	How connected do you feel to your ethnic community?
\item	How does your family view STEM subjects related subjects?
\item	Have you ever taken a computer science or coding class as an extracurricular activity
\item	How often do you use a computer, a laptop, an iPad or a phone for school work?
\item	How often do you use computers or other digital devices for personal activities?
\item	What kind of job would you like to have when you grow up?
\item	Can you think of any women who are really good at computers?
\item	Is computer science considered geeky by your peers?
\end{itemize}

\paragraph{Perceptions of CS}
This section included a short answer question: "What comes to mind when you think about computer science?" and rating questions below.
\begin{itemize}
\item	 I think computer science is enjoyable
 \item	I am interested in studying computer science
 \item	I want to study computer science because it pays well
 \item	I feel good about learning computer science at school
 \item	I am confident in my ability to do well in computer science courses
 \item	I think computer science is for boys only
 \item	I think that boys and girls can be good at computer science
 \item	I want to have a job someday that involves computers
 \item	I think anyone can learn computer science
 \item	I think that learning computer science is hard
 \item	I think you need to like competition to study computer science
People who do computer science need to be very smart
\end{itemize}

\paragraph{Collaboration and Teamwork}
\begin{itemize}
\item I feel good about working with my team mates on projects
\item	I think it is important for us to work together when we have tasks to do
\item	I like when we all help each other to finish our work
\item	It’s easy for me to talk to my teammates and share my ideas when we’re working on something
\item	I am okay with doing different jobs in our group when we need to
\item	I think it is cool when we have all sorts of people in our group
\item   I think boys and girls can work together well in a group
\item	I would not mind being the only girl in the group
\item	I do not think I can work well if I am the only girl in the room
\item	I think that a computer science club is a boy’s club
\item	I do not think I work well in a group
\item	I like working alone more than working in a team
\end{itemize}

\paragraph{Relating to Computer Science in the Real World}
\begin{itemize}
\item We can use computer science to solve real-world problems
\item	You need to be a scientist or engineer to use computer science for your work
\item	Computer Science can help us learn and explore new things.
\item	People can use computer science to find solutions to environmental problems
\item	Computer science relates to what I experience in the real world
\item	I think that learning computer science can help me solve real-world problems
\item	I believe that computer science can help solve global challenges like poverty
\end{itemize}

\section{Post-workshop Questionnaire}
\label{appendix:postsurvey}

\paragraph{Active Perception of CS}
\begin{itemize}
\item During the tasks, I felt confident when using a computer.
\item During the tasks, I found computer science enjoyable.
\item During the tasks, I found it interesting to study computer science.
\item After doing the tasks, I am feeling good about studying computer science.
\item After doing the tasks, I am feeling confidant in my ability to do well in computer science.
\item After doing the tasks, I think computer science is for boys only.
\item After doing the tasks, I think that boys and girls can be good at computer science.
\item After doing the tasks, I want to have a job someday that involves computers.
\item While doing the tasks, I felt that anyone can learn computer science.
\item While doing the tasks, I felt learning computer science is hard.
\item While doing the tasks, I felt that people who do computer science need to be very smart.
\end{itemize}

\paragraph{Attitudes towards Collaboration}
\begin{itemize}
    \item During the tasks, I felt good about working with my teammate.
    \item During the tasks, I found important for us to work together when we were doing the tasks.
    \item During the tasks, I liked when we all helped each other to finish our work.
    \item While doing the tasks, it was easy for me to talk to my teammates and share my ideas when we’re working on something
    \item During the workshop, I was doing different jobs in our group when we need to.
    \item I think it’s cool when we have all sorts of different people in our team.
    \item After doing the workshop, I would not mind being the only girl in a group.
    \item After doing the workshop, I do not think I can work well if I am the only girl in a group.
    \item I do not think I worked well in my group today.
    \item I would have preferred to work alone over working in a team.
    \item It was easy for me to talk to my teammates and share my ideas during the tasks.
    \item I think we could have benefited from having a boy in our team.
\end{itemize}

\paragraph{The Effect of Exposure on Relating Computer Science to Real World}
\begin{itemize}
    \item While doing the tasks, I found computer science useful to solve real-world problems. 
    \item After doing the workshop today, I think you need to be a scientist or engineer to use computer science for your work.
    \item While doing the tasks, I found that computer science can help us learn and explore new things
    \item After doing the tasks, I now think people can use computer science to find solutions to environmental problems. 
    \item After doing the tasks, I found that computer science relates to things I experience in the real world. 
    \item After doing the tasks, I think that learning computer science can help me solve real-world problems.  
    \item After doing the tasks, I believe that computer science can help solve global challenges.
    \item During the last task, I liked that we could come up with our own ideas.  
    \item During the last task, I found it easy to think about how computer science can be used in our industry. 
\end{itemize}

\paragraph{Meaningful Connections} 
\begin{itemize}
    \item I cared about the problem in the tasks, so I wanted to solve it more.  
    \item I think computers can help make the world a better place by solving problems like pollution.  
    \item Working on a computer project that helps make the world better would make me really happy.  
    \item Knowing I'm helping the world with my computer skills would make me try even harder on my tasks.  
    \item Working on a project that helps the world using computers makes me feel really powerful.  
    \item The workshop idea built on United Nations Sustainable Development Goals was meaningful and relevant.  
    \item After today's workshop, I feel like I might want to have a job where I use computers to help the Earth. 
\end{itemize}

\paragraph{Reflections on Careers} 
\begin{itemize}
    \item After doing today’s workshop, I think that computer science is important for most jobs. 
    \item After today’s workshop, I would like to learn more about careers in computer science. 
    \item After today’s workshop, I think that computer science is part of all jobs.  
    \item After today’s workshop, I think that having knowledge of computer science will open more job opportunities
    \item Doing today's tasks about real-world issues made me think more about working in computer science.  
    \item Working on today’s tasks helped me understand what kinds of job I could have in the future. 
    \item Doing today’s tasks about everyday stuff with computers made me feel like I could be good at tech jobs. 
    \item Solving problems with computers made me feel like I could be really creative and make new things. 
    \item Learning how computers can solve everyday problems made me excited about learning more about tech jobs.
    \item Doing projects with computers that help with real-life things made me think that being a computer scientist could be a cool job
\end{itemize}

\section{Generational Status \& Pupils’ Preconceptions}
\label{appendix:GenerationalStatus}
\begin{figure}[h!]
\centering
\includegraphics[width=1\linewidth]{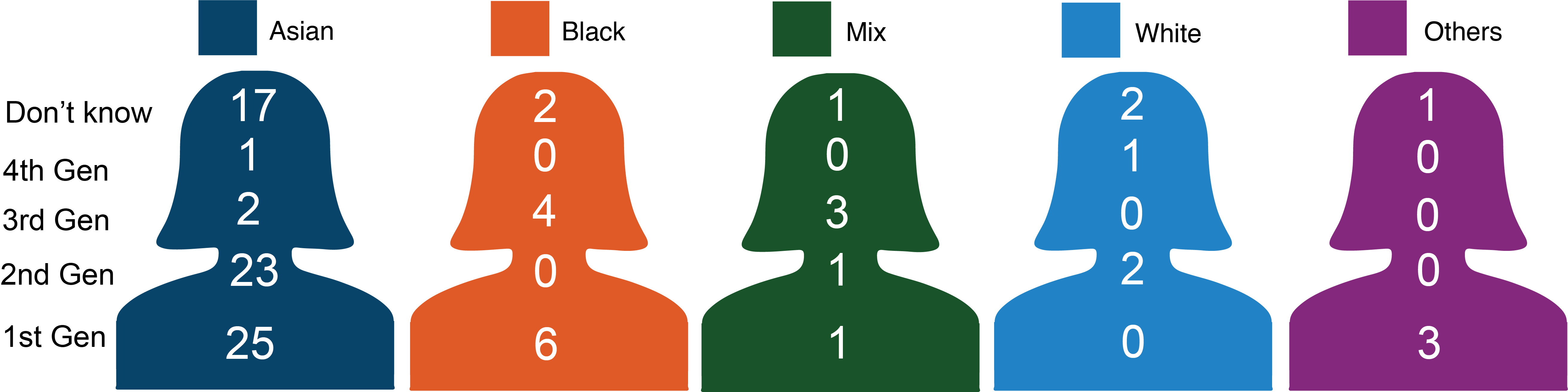}
\caption{Distribution of immigration generational status across ethnic groups. Counts are shown for each generation, with categories colour-coded by ethnicity.}
\label{fig:GenStatus}
\end{figure}

In addition to ethnicity, we examined whether immigration generational status influenced pupils’ preconceptions of computer science, teamwork, and collaboration. This was based on the assumption that the longer a family has been settled in the UK across generations, the more positively pupils may perceive computer science \cite{Kennedy2010}. See Figure~\ref{fig:GenStatus} for the distribution of generational status by ethnicity. 

We conducted the Kruskal--Wallis test, which highlights generational differences in both preconceptions of Computer Science and attitudes toward collaboration, communication, and teamwork. First-generation pupils tended to feel less positive about collaboration on projects, whereas second- and third-generation pupils expressed higher confidence in working with others ($H(4) = 17.34$, $p = 0.001$). Second-generation pupils in particular had significant results with respect to working in teams, while first-generation and ``do not know'' respondents were less strongly aligned with this view ($H(4) = 14.32$, $p = 0.006$). Third-generation pupils found it easier to talk to teammates compared to first-generation peers ($H(4) = 13.9$, $p = 0.007$). In terms of preconceptions of CS, first-generation pupils were more likely to agree with the notion that ``CS is a boys' club'', while later generations were less likely to endorse this stereotype ($H(4) = 10.32$, $p = 0.035$). Post-hoc comparisons showed that these effects were driven by pupils who reported not knowing their immigrant generation, who differed significantly from second- and fourth+-generation pupils in their views on collaboration and teamwork.

\end{document}